\documentclass{elsarticle}

\usepackage{amssymb,amsmath,amsthm}
\usepackage{xspace}
\usepackage{longtable}

\theoremstyle{plain}
 \newtheorem{theorem}{Theorem}
 \newtheorem{lemma}[theorem]{Lemma}
 \newtheorem{proposition}[theorem]{Proposition}
 
 \newtheorem{corollary}[theorem]{Corollary}
\theoremstyle{definition}

\theoremstyle{remark}

\theoremstyle{remark}

\begin{document}

\begin{frontmatter}

\title{On primary and secondary repetitions in words}

\author[msu]{Roman Kolpakov}
\ead{foroman@mail.ru}

\address[msu]{Moscow State University,
Leninskie Gory, 119992 Moscow, Russia}

\begin{abstract}
Combinatorial properties of maximal repetitions (runs)
in formal words are studied. We classify all maximal 
repetitions in a word as primary and secondary where 
the set of all primary repetitions determines all the
other repetitons in the word. Essential combinatorial
properties of primary repetitions are established.
\end{abstract}

\end{frontmatter}

\section{Inroduction}

Let $w=w[1]w[2]\ldots w[n]$ be an arbitrary word. The length~$n$
of~$w$ and is denoted by $|w|$. A word $w[i]\cdots w[j]$, where $1\le 
i\le j\le n$, is called a {\it factor} of~$w$ and is denoted by $w[i..j]$.
Note that factors can be considered as fragments of the original word or
as words themselves. So for factors we have two different notions of equality:
factors can be equal as the same fragment of the original word or as the 
same word. To avoid this ambiguity, we will use two different notations:
if two factors $u$ and $v$ are the same word (the same fragment of the original 
word) we will write $u=v$ ($u\equiv v$). For any $i=1,\ldots,n$ the factor 
$w[1..i]$ ($w[i..n]$) is called a {\it prefix} (a {\it suffix}) of~$w$.
A positive integer $p$ is called  a {\it period} of~$w$ if $w[i]=w[i+p]$ 
for each $i=1,\ldots ,n-p$. We denote by $p(w)$ the minimal period of~$w$
and by $e(w)$ the ratio $|w|/p(w)$ which is called the {\it exponent} of~$w$.
A word is called {\it primitive} if its exponent is not an integer greater
than~1.

By repetition in a word we mean any factor of exponent greater than or equal to~2.
Repetitions are fundamental objects, due to their primary importance in word 
combinatorics~\cite{Lothaire83} as well as in various applications, such as 
string matching algorithms~\cite{GaliSeiferas83,CrochRytter95}, molecular 
biology~\cite{Gusfield97}, or text compression~\cite{Storer88}.
The simplest and best known example of repetitions is factors of the form $uu$, 
where $u$ is a nonempty word. Such repetitions are called {\it squares}.
We will call the first (second) factor~$u$ of the square $uu$ {\it the left (right)
root} of this square. Avoiding ambiguity\footnote{Note that the period of a square 
is not necessarily the minimal period of this word.}, by the {\it period} of a square 
we will mean the length of its roots. A square is called {\it primitive} if its roots
are primitive. Primitive squares are a particular case of factors of the form 
$u^k=\underbrace{uu\ldots u}_k$ where $k>1$ and $u$ is a nonempty primitive word.
Such factor is called a {\it primitive integer power} with the root~$u$.
A primitive integer power is called {\it maximal} if it cannot be extended
to the left or to the right in the word by at least one root. Note that any
primitive integer power is contained in only one maximal integer power.
In an analogous way, one can note that any repetition is contained
in only one {\it maximal} repetition with the same minimal period
which cannot be extended to the left or to the right in the word by 
at least one letter with preserving its minimal period. Maximal repetitions
are usually called {\it runs} in the literature. Since runs contain all the other
repetitions in a word, the set of all runs can be considered as a compact encoding 
of all repetitions in the word which has many useful applications (see, 
for example,~\cite{Crochetal1}).

Questions concerning the maximum possible number of repetitions in words are
actively investigated in the literature. In particular, it is shown 
in~\cite{Crochemor81,CrochRytter95} that the maximum possible number
of primitive square and maximal integer powers in words of length~$n$
is $\Theta (n\log n)$. It is proved in~\cite{KK00} that, unlike the case
of maximal integer powers, the maximum possible number $mrn(n)$ of runs
in words of length~$n$ is $O(n)$ and, moreover, the maximum possible sum
$mex(n)$ of all runs in words of length~$n$ is also $O(n)$. Due to a series
of papers~\cite{Rytter06,Rytter07,PugSimSmyth,CrochIlie07,Giraud,CrochIlie08,CrochIlieTinta}
more precise upper bounds on $mrn(n)$ have been obtained. For the present time
the best upper bound $1.029 n$ on $mrn(n)$ is obtained in~\cite{CrochIlieTinta}.
The problem of low bounds on $mrn(n)$ is considered in~\cite{FraSimSmyth,FraYang,Simpson}.
More precise bounds on $mex(n)$ have been also obtained in~\cite{CrochIlie07,CrochIlie08,Crochetal11}.
In particular, the best known bounds $mex(n)\le 4.1 n$ and $mex(n)>2.035 n$ are obtained 
in~\cite{Crochetal11}. Analogical estimates for runs with exponent at least~3 are obtained 
in~\cite{Crochetal10,Crochetal11}.

Further we denote by $R (w)$ the set of all maximal repetitions in a word~$w$.
Let $\lambda$ be a natural number. For maximal repetitions, in our opinion, 
one could make the two following natural conjectures:
\begin{enumerate}
\item  The number of maximal repetitions with the minimal period not less than~$\lambda$ 
in the word~$w$ is upper bounded by $\varphi (\lambda) n$ where $\varphi (\lambda)\to 0$
as $\lambda\to\infty$.
\label{fstconj}
\item The maximal number of maximal repetitions containing the same letter~\footnote{By the same
letter we mean that letters in different positions of the word are different.} in the word~$w$ 
is $o(n)$.
\label{scdconj}
\end{enumerate}
Unfortunately, both the conjectures are not true. As a counterexample, we can consider
the word $w_k=(01)^k(10)^k$ of length $4k$. It is easy to check that
$R (w_k)=\{r_1, r_2,\ldots, r_{k+2}\}$ where $r_1=(01)^k$, $r_2=(10)^k$, and
$r_i=(1(01)^{k-3})^2$ for $i=3, 4,\ldots, k+2$. Thus, for any $\lambda >2$
the word $w_{\lambda}$ contains $\lfloor\lambda /2\rfloor=\Omega (|w_{\lambda}|)$
maximal repetitions with the minimal period not less than~$\lambda$ which
contradicts conjecture~\ref{fstconj}. Moreover, the middle letters of $w_k$
are contained in $k+1=\Omega (|w_k|)$ different maximal repetitions from $R (w_k)$
which contradicts conjecture~\ref{scdconj}. However, one can easily observe that
$w_k$ has actually two ``original'' adjacent maximal repetitions $r_1$ and $r_2$
which ``generate'' all the other repetitions $r_3, r_4,\ldots, r_{k+2}$. This
observation suggests that it would be possible to indicate in $R (w)$ a subset
of repetitions which ``generate'' all the other maximal repetitions of~$w$.
In this paper we formally define the notion of generation of repetitions.
In accordance with this notion, generated repetitions are called secondary
and all the other maximal repetitions are called primary. Originally the notions of 
primary and secondary repetitions were introduced in~\cite{GasKolPot} where
they was used for space efficient search for maximal repetitions. In~\cite{GasKolPot}
some auxiliary combinatorial results for primary and secondary repetitions
are also obtained. The notions of primary and secondary repetitions defined here
are slightly different from the notions introduced in~\cite{GasKolPot}. However,
this difference is not crucial. Thus, in the present paper we continue the combinatorial
investigations started in~\cite{GasKolPot} for primary and secondary repetitions.
In particular, we show that, unlike the case of all maximal repetitions, both
conjectures~\ref{fstconj} and \ref{scdconj} are true for primary repetitions.
More precisely, we prove that in the word~$w$ the sum of exponents of all
primary repetitions with the minimal period not less than~$\lambda$ and 
all secondary repetitions generated by these primary repetitions is $O(n/\lambda)$
which obviously implies that the number of primary repetitions with the minimal period 
not less than~$\lambda$ in the word~$w$ is also $O(n/\lambda)$. Moreover, we prove
that the maximal number of primary repetitions which have the minimal period not 
less than~$\lambda$ and contain the same letter in the word~$w$ is $O(\log (n/\lambda))$
which obviously implies that conjecture~\ref{scdconj} is also true for primary repetitions.
Thus, the set of all primary repetitions which represent actually all repetitions 
in a word is more convenient for considering and treatment than the set of all
maximal repetitions. 

\section{Auxiliary definitions and results}

The results of the paper are based on the following well-known fact
which is usually called {\it the periodicity lemma}.
\begin{lemma}
If a word $w$ has two periods $p,q$, and $|w|\ge p+q$,
then $\gcd (p,q)$ is also a period of~$w$.
\label{perilemma}
\end{lemma}

Using the periodicity lemma, it is easy to obtain
\begin{proposition}
Let $q$ be a period of a word~$w$ such that $|w|\ge 2q$.
Then $q$ is divisible by $p(w)$.
\label{repper}
\end{proposition}

We will use also the following evident fact.
\begin{proposition}
If two factors of a word have the same period~$q$ and are 
overlapped by at least $q$ letters then $q$ is a period
of the union of these factors.
\label{unionper}
\end{proposition}

Let $w=w[1]w[2]\ldots w[n]$ be an arbitrary word.
A repetition $r\equiv w[i..j]$ in~$w$ is called {\it maximal}
if it satisfies the following conditions:
\begin{enumerate}
\item if $i>1$, then $w[i-1]\neq w[i-1+p(r)]$,
\item if $j<n$, then $w[j+1-p(r)]\neq w[j+1]$.
\end{enumerate}
In other words, a repetition in~$w$ with the minimal 
period $p$ is maximal if its one letter extension in~$w$ 
(to the left or to the right) results in a factor with
the minimal period $>p$. It is obvious that any repetition
in a word is contained in only one maximal repetition with
the same minimal period. We denote by $R (w)$ the set of 
all maximal repetitions in~$w$.
The following fact about maximal repetitions is a trivial
consequence of Proposition~\ref{unionper}.
\begin{proposition}
The overlap of two non-separated different maximal repetitions 
with the same minimal period~$p$ is smaller than~$p$.
\label{overlap1}
\end{proposition}

Proposition~\ref{overlap1} obviously implies
\begin{proposition}
Let $r'\equiv w[i'..j']$, $r''\equiv w[i''..j'']$, $r'''\equiv w[i'''..j''']$
be different maximal repetitions in~$w$ with the same minimal period and
$i'\le i''\le i'''$. Then $r'$ and $r'''$ are not non-separated.
\label{nonoverlap}
\end{proposition}

\section{Primary and secondary repetitions}

Let $r$ be a repetition in the word~$w$. We call any factor 
of~$w$ which has the length $p(r)$ and is contained in~$r$ a 
{\it cyclic root} of~$r$. Note that for any cyclic root $u$ of~$r$ 
the word~$r$ is a factor of the word $u^k$ where $k$ is a large 
enough number. So it follows from the minimality of the period $p(r)$ 
that any cyclic root of~$r$ has to be a primitive word. Hence any two 
adjacent cyclic roots of~$r$ form a primitive square with the period $p(r)$ 
which is called a {\it cyclic square} of~$r$. Two repetitions $r'$ and $r''$ 
with the same minimal period~$p$ are called {\it cognate} if the words
$r'$ and $r''$ are factors of the same word $u^k$, where $|u|=p$ and $k$ 
is a large enough number. It easy to see that cognate repetitions have
the same set of distinct cyclic roots. For cognate repetitions we have the
following statement which is proved in~(\cite{GasKolPot}, Lemma~1).

\begin{lemma}
Let $r', r''$ be cognate repetitions with minimal period~$p$ in the word~$w$. 
Then for any cyclic roots $u'\equiv w[i'_u..i'_u+p-1]$, $v'\equiv w[i'_v..i'_v+p-1]$ 
of~$r'$ and any cyclic roots $u''\equiv w[i''_u..i''_u+p-1]$, $v''\equiv w[i''_v..i''_v+p-1]$ 
of~$r''$ such that $u'=u''$, $v'=v''$ an equality $i''_u-i'_u\equiv i''_v-i'_v\pmod{p}$ 
holds.
\label{repsamroot}
\end{lemma}

Lemma~\ref{repsamroot} implies that there exists a residue class modulo~$p$, 
such that, for any equal cyclic roots $u'\equiv w[i'_u..i'_u+p-1]$
of~$r'$ and $u''\equiv w[i''_u..i''_u+p-1]$ of~$r''$, the value $i''-i'$ 
belongs to this class. We denote by $\sigma (r',r'')$ the minimal non-negative 
residue of this class. 
It is easy to see that cognate non-separated repetitions $r',r''$ are extended 
to the same maximal repetition if $\sigma (r',r'')=0$. Hence

\begin{proposition}
For any different cognate non-separated maximal repetitions $r',r''$ the value 
$\sigma (r',r'')$ is positive.
\label{sisposit}
\end{proposition}

We use also the following fact which is proved actually in~(\cite{GasKolPot}, Lemma~2)
(here we present a shorter proof of this fact).

\begin{lemma}
Let $r'\equiv w[i'..j']$, $r''\equiv w[i''..j'']$ be cognate non-separated repetitions with minimal 
period~$p$, and $v\equiv w[l..l+2q-1]$ be a primitive square with the period~$q$ such that $q\ge 2p$
and $v$ is contained completely in $w[i'..j'']$. Than $i''\le l+q\le j'+1$.
\label{cycsquar}
\end{lemma}

{\bf Proof.} Denote respectively the roots $w[l..l+q-1]$ and $w[l+q..l+2q-1]$ of~$v$ by $u'$ and $u''$.
Suppose that $l+q<i''$. Note that in this case $u'$ is contained completely in $r'$, 
so $p$ is a period of~$u'$. Therefore, $p$ is also a period of~$u''$. If $v$ is contained 
completely in $r'$ than $v$ has both periods $p$ and $q$ such that $|v|=2q>p+q$.
So by the periodicity lemma in this case $v$ has also the period $\gcd (p, q)$
which contradicts the primitivity of roots of~$v$. Thus we can suppose that $l+2q-1>j'$.
Let $j'+1-p\ge l+q$. Than both letters $w[j'+1-p]$ and $w[j'+1]$ are contained in $u''$. 
Since $p$ is a period of~$u''$, we obtain that $w[j'+1-p]=w[j'+1]$ which contradicts
that $r'$ is maximal. Now let $j'+1-p<l+q$. Taking into account that $j'\ge i''-1$,
in this case we have
$$
l+2q-1>j'-p+q\ge j'+p\ge i''+p-1>i''-1\ge l+q,
$$
so both letters $w[i''+p-1]$ and $w[i''-1]$ are contained in $u''$.
Since $p$ is a period of~$u''$, we conclude that $w[i''+p-1]=w[i''-1]$ which contradicts
that $r''$ is maximal. Thus $i''\le l+q$. The inequality $l+q\le j'+1$ is proved by
symmetrical way.

Let $r'\equiv w[i'..j']$, $r''\equiv w[i''..j'']$ where $i'\le i''$ be cognate non-separated 
repetitions from $R (w)$ with minimal period~$p$. Then it follows from Proposition~\ref{overlap1}
that $i'<j'+1-p<i''\le j'+1<j''$. We say that a repetition $r\equiv w[i..j]$ from $R (w)$ {\it is
generated} by repetitions $r'$ and $r''$ if the following conditions are valid:
\begin{enumerate}
\item $p(r)\ge 3p$;
\item $i'< i\le j'$,
\item $i''\le j < j''$.
\end{enumerate}
We will also say in this case that $r'$ ($r''$) generates~$r$ from left (from right).
If a repetition is generated by some repetitions from from $R (w)$ we call this
repetition {\it secondary}. All repetitions from $R (w)$ which are not secondary
are called {\it primary}. By $Rp (w)$ we denote the set of all primary repetitions 
in~$w$, and by $Rs (w)$ we denote the set of all secondary repetitions in~$w$.

\begin{lemma}
Any secondary repetition is generated by only one pair of repetitions.
\label{paruniq}
\end{lemma}

\noindent
{\bf Proof:} Let a maximal repetition $r$ be generated by a pair 
$(r'_1, r''_1)$ of repetitions with a minimal period $p_1$ and a pair 
$(r'_2, r''_2)$ of repetitions with a minimal period $p_2$ where
$r'_k\equiv w[i'_k..j'_k]$, $r''_k\equiv w[i''_k..j''_k]$ for $k=1, 2$.
Consider in~$r$ an arbitrary cyclic square $v\equiv w[l..l+2p(r)-1]$. 
Since $v$ is contained completely in $w[i'_1..j''_1]$ and $w[i'_2..j''_2]$,
by Lemma~\ref{cycsquar}, we have $i''_k\le l+p(r)\le j'_k+1$
for $k=1, 2$. Therefore, the left root $w[l..l+p(r)-1]$ of~$v$ is contained
in both repetitions $r'_1$ and $r'_2$. So $r'_1$ and $r'_2$ are
overlapped by at least $p(r)$ letters where $p(r)>p_1+p_2$. 
Moreover, the right root $w[l+p(r)..l+2p(r)-1]$ of~$v$ is contained
in both repetitions $r''_1$ and $r''_2$. So $r''_1$ and $r''_2$ are
also overlapped by at least $p(r)$ letters. Hence, by Proposition~\ref{overlap1}, 
we have $r'_1\equiv r'_2$ and $r''_1\equiv r''_2$.

On the other hand, we can describe explicitly all repetitions
generated by a given pair of repetitions.

\begin{lemma}
Let a maximal repetition $r$ in a word~$w$ be generated by a 
pair $(r', r'')$ of repetitions with a minimal period $p$ 
where $r'\equiv w[i'..j']$, $r''\equiv w[i''..j'']$. Then 
$p(r)=\alpha p+\sigma (r',r'')$, and $r\equiv w[i''-p(r)..j'+p(r)]$, where
$\alpha$ is an arbitrary integer satisfying the inequalities
\begin{equation}
3\le\alpha <\frac{1}{p}\bigl(\min\{i''-i',j''-j'\}-
\sigma (r',r'')\bigr).
\label{condalpha}
\end{equation}
\label{genereps}
\end{lemma}

{\bf Proof:} Consider in~$r$ an arbitrary cyclic square $v\equiv w[l..l+2p(r)-1]$.
Let $u'\equiv w[l..l+p-1]$ ($u''\equiv w[l+p(r)..l+p(r)+p-1$) be the prefix of 
length~$p$ in the left (right) root of~$v$. By Lemma~\ref{cycsquar} we have
$i''\le l+p(r)\le j'+1$, so the left root of~$v$ is contained in~$r'$ and 
the right root of~$v$ is contained in~$r''$. Hence $u'$ is a cyclic root of~$r'$ and 
$u''$ is a cyclic root of~$r''$. Since $u'=u''$, by Lemma~\ref{repsamroot} we obtain 
that $l+p(r)-l=p(r)\equiv \sigma (r',r'')$, i.e. $p(r)=\alpha p+\sigma (r',r'')$.
Since $\sigma (r',r'')<p$, from $p(r)\ge 3p$ we have $\alpha\ge 3$. 
If $\alpha$ satisfies inequalities~(\ref{condalpha}), then it is easy
to note that all factors $w[l..l+2p(r)-1]$ such that $i''\le l+p(r)\le j'+1$
are cyclic square of the same maximal repetition $w[i''-p(r)..j'+p(r)]$ which
is generated by $(r', r'')$. Thus $r\equiv w[i''-p(r)..j'+p(r)]$. Using Lemma~\ref{cycsquar},
it is also not difficult to see that if $\alpha\ge\frac{1}{p}\bigl(\min\{i''-i',j''-j'\}-\sigma (r',r'')\bigr)$,
either $w[i'..j'']$ doesn't contain primitive squares with the period $\alpha p+\sigma (r',r'')$
or such squares are cyclic square of a repetition $w[i..j]$ which doesn't satisfy
the conditions $i'<i$ or $j<j''$. 

\begin{corollary}
Any secondary repetition is generated by a pair of primary
repetitions.
\label{secbyprim}
\end{corollary}

{\bf Proof:} Let $r=w[i..j]$ be a secondary repetition generated by a pair 
$(r',r'')$ of repetitions with minimal period~$p$ in a word~$w$ where 
$r'\equiv w[i'..j']$, $r''\equiv w[i''..j'']$. Then, by Lemma~\ref{genereps}, 
we have 
\begin{equation}
e(r)=\frac{2p(r)+\delta}{p(r)}=2+\frac{\delta}{p(r)}
\label{forer}
\end{equation}
where $\delta$ is the overlap of repetitions $r',r''$.
Since $p(r)\ge 3p$ and $\delta<p$, due to Proposition~\ref{overlap1},
the equality~(\ref{forer}) implies $e(r)<7/3$. Thus the exponent
of any secondary repetition is less than~$7/3$. Consider now
the repetition $r'=w[i'..j']$. Since $i=i''-p(r)$ by Lemma~\ref{genereps},
we have
$$
|r'|\ge i''-i'>i''-i=p(r)\ge 3p.
$$ 
Hence $e(r')>7/3$. Similarly we can prove that $e(r'')>7/3$. So neither $r'$ 
nor $r''$ can be a secondary repetition. 

Using Lemma~\ref{condalpha} and Corollary~ref{secbyprim}, we can easily 
compute all secondary repetitions from the set of all primary repetitions.
So the set $Rp (w)$ represents actually all repetitions in~$w$.

\begin{corollary}
Any repetition~$r$ generates from left less than $e(r)-2$ repetitions.
\label{descr}
\end{corollary}

{\bf Proof:} It is easy to see from Proposition~\ref{nonoverlap} that any maximal repetition~$r$ 
can have to the right only one maximal repetition $r'$ non-separated and cognate with~$r$.
Thus all repetitions generated by~$r$ from left have to be generated by only one pair
$(r, r')$ of repetitions. From Lemma~\ref{genereps} we conclude that the number of repetitions 
generated by this pair is no more than the number of integer~$\alpha$ such that $3\le\alpha < e(r)$
which is obviously less than $e(r)-2$.

\section{Main results}

Further we consider pairs of integers $(p, j)$ where $p>0$. We will call 
such pairs {\it points}.  For any two points $(p', j')$, $(p'', j'')$ we say 
that the point $(p', j')$ {\it covers} the point $(p'', j'')$ if $p'\le p'' \le 4p'/3$ 
and $j'-(2p'/3)\le j''\le j'$. By $V(p, j)$ we denote the set of all points 
covered by the point $(p, j)$. Let ${\cal E}(w)$ be the set of all points $(p, j)$ 
such that $1\le p\le 2n/3$ and $1\le j\le n$. For any repetition $r\equiv w[i..j]$ 
from $R(w)$ we denote by ${\cal P}(r)$ the set of all points $(p(r), i+kp-1)$ 
of ${\cal E}(w)$ where $k$ is an integer greater than or equal to~2  and $i+kp-1\le j$. 
Note that $|{\cal P}(r)|=\lfloor e(r)-1\rfloor$, so for any repetition~$r$ 
the set ${\cal P}(r)$ is not empty. Moreover, from Proposition~\ref{overlap1} we have

\begin{proposition}
For any different repetitions $r'$, $r''$ from $R(w)$ the sets ${\cal P}(r')$ and 
${\cal P}(r'')$ are not intersected. 
\label{notinter}
\end{proposition}

We also use the following fact.

\begin{proposition}
Two different points $(p', j')$, $(p'', j'')$ of ${\cal E}(w)$ such that $p'=p''$
can not cover the same point.
\label{notcover}
\end{proposition}

{\bf Proof.} Let $p'=p''$. Then $j'\neq j''$. Assume without loss of generality that $j''<j'$.
Let the points $(p', j')$, $(p'', j'')$ cover the same point $(p, j)$. Then
$j'-(2p'/3)\le j\le j''<j'$. So the points $(p', j')$, $(p'', j'')$ can not
be contained in the same set ${\cal P}(r)$. On the other hand, if $(p', j')$ and
$(p'', j'')$ are contained in the sets ${\cal P}(r')$, ${\cal P}(r'')$ for some
different repetitions $r'$ and $r''$ with the same minimal period~$p'=p''$ then
these repetitions have an overlap of length greater than or equal to $4p'/3$
which contradicts Proposition~\ref{overlap1}.

By ${\cal E}'(w)$ we denote the subset $\bigcup_{r\in Rp (w)} {\cal P}(r)$ of ${\cal E}(w)$.
Note that, by Proposition~\ref{notinter}, each point of ${\cal E}'(w)$ belongs
to only one set ${\cal P}(r)$.

Our results are based on the following statement.

\begin{lemma}
Three different points of ${\cal E}'(w)$ can not cover the same point.
\label{baselemma}
\end{lemma}

{\bf Proof.} Let three different points $(p_1, j_1)$, $(p_2, j_2)$ and $(p_3, j_3)$ 
of ${\cal E}'(w)$ cover the same point $(p, j)$. Then, by Proposition~\ref{notcover},
the numbers $p_1$, $p_2$ and $p_3$ have to be pairwise different. Assume without loss 
of generality that $p_3<p_2<p_1$. Note that in this case we have
$$
p_3<p_2<p_1\le p\le 4p_3/3.
$$
For $k=1,2,3$ let $r_k=w[s_k..t_k]$ be the primary repetition such that $(p_k, j_k)\in{\cal P}(r_k)$.
Note that $p(r_k)=p_k$. Denote $j_k-2p_k+1$ by $i_k$. Note that the factor $w[i_k..j_k]$ 
is contained completely in $r_k$, so in $r_k$ we can consider the conjugate cyclic roots 
$w[i_k..j_k-p_k]$ and $w[i_k+p_k..j_k]$. Denote respectively these roots by $u'_k$ and $u''_k$. 
We also denote $p_2-p_3$ by~$q$. From $p_2<4p_3/3$ we have $q<p_3/3$. To prove the lemma, 
we consider separately the three following cases.

Case I. Let $j_2\le j_3$. Note that in this case $i_2<i_3$. First we prove that 
in this case $s_3=i_3$, i.e. $r_3$ can not be extended with the same period to the left 
of $w[i_3]$. Assume that $s_3\neq i_3$. Then, by definiton of ${\cal P}(r_3)$, 
the repetition $r_3$ has at least one cyclic root to the left of $w[i_3]$, i.e. the factor 
$w[i_3-p_3..j_3]$ is contained completely in $r_k$. So $p_3$ is a period of $w[i_3-p_3..j_3]$. 
Let $i_3-p_3\le i_2$. Then the factor $w[i_2..j_2]$ is contained in $w[i_3-p_3..j_3]$.
So $w[i_2..j_2]$ has both periods $p_2$ and $p_3$. Moreover, $|w[i_2..j_2]|=2p_2>p_2+p_3$.
Therefore, by the periodicity lemma $w[i_2..j_2]$ has the period $\gcd (p_2, p_3)$
which contradicts the primitivity of cyclic roots of~$r_2$. Now let $i_3-p_3>i_2$.
Then the overlap $w[i_3-p_3..j_2]$ of factors $w[i_3-p_3..j_3]$ and $w[i_2..j_2]$
has both periods $p_2$ and $p_3$. Since $j_3-(2p_3/3)\le j\le j_2\le j_3$ and $p_2<4p_3/3$,
we have
$$
|w[i_3-p_3..j_2]|=3p_3-(j_3-j_2)\ge 7p_3/3>p_2+p_3.
$$
Therefore, by the periodicity lemma $w[i_3-p_3..j_2]$ has the period $\gcd (p_2, p_3)$
which contradicts again the primitivity of cyclic roots of~$r_2$. Thus, $s_3=i_3$. Since $i_3>1$, 
it implies that $w[i_3-1]\neq w[i_3+p_3-1]$. It is easy to see that $w[i_3+p_3-1]$ 
is contained in $u''_2$. So from $u'_2=u''_2$ we have $w[i_3+p_3-1]=w[i_3+p_3-1-p_2]=w[i_3-q-1]$. 
Thus $w[i_3-1]\neq w[i_3-q-1]$. Denote by~$v$ the overlap $w[i_3+p_3..j_2]$ of $u''_2$ 
and $u''_3$. Taking into account $j_2\ge j\ge j_3-(2p_3/3)$, we obtain
$$
|v|=j_2-(j_3-p_3)\ge\frac{p_3}{3}>q.
$$
Moreover, since $u'_2=u''_2$ and $u'_3=u''_3$, we have 
\begin{equation}
v=w[i_3..j_2-p_3]=w[i_3+p_3-p_2..j_2-p_2]
\label{doublev}
\end{equation}
which implies that $q$ is a period of~$v$. For case I we consider separately subcases 
$i_1<i_3-q$ and $i_1\ge i_3-q$.

Subcase I.1. Let $i_1<i_3-q$. Since $j_3-(2p_3/3)\le j\le j_1$ and $p_1\le 4p_3/3$,
the relation $j_1-p_1\ge j_3-2p_3=i_3-1$ is valis. Thus in this case we have
$i_1\le i_3-q-1 < 1_3-1\le j_1-p_1$. So both symbols $w[i_3-q-1]$ and $w[i_3-1]$
are contained in $u'_1$. Since $u'_1=u''_1$, we obtain $w[i_3-q-1]=w[i_3+p_1-q-1]$
and $w[i_3-1]=w[i_3+p_1-1]$. Therefore, $w[i_3+p_1-q-1]\neq w[i_3+p_1-1]$.
Using the inequalities $p_1>p_2$ è $p_1\le 4p_3/3$, we obtain $i_3+p_1-q-1\ge 
i_3+p_2-q=i_3+p_3$ and $i_3+p_1-1=j_3-2p_3+p_1\le j_3-(2p_3/3)\le j\le j_2$.
Thus both symbols $w[i_3+p_1-q-1]$ and $w[i_3+p_1-1]$ are contained in~$v$.
So $w[i_3+p_1-q-1]\neq w[i_3+p_1-1]$ contradicts the fact that $q$ is a period of~$v$.
So this subcase is impossible.

Subcase I.2. Let $i_1\ge i_3-q$. Note that in this case $j_3<j_1$. Consider the factor 
$v'\equiv w[i_3-q..j_2-p_3]$. Note from~(\ref{doublev}) that $q$ is a period of~$v'$. 
Moreover, $|v'|=|v|+q>2q$. So $v'$ is a repetition. Let $q'$ be the minimal period of~$v'$. 
Note that $q'$ is a divisor of~$q$ by Proposition~\ref{repper}. Let $\hat v'\equiv w[i'..j']$
be the maximal repetition containing~$v'$. Then we consider separately subcases
$j'<j_3-p_3$ and $j'\ge j_3-p_3$.

Subcase I.2.a. Let $j'<j_3-p_3$. Since the repetition $\hat v'$ is maximal, we have
$w[j'+1]\neq w[j'+1-q']$. It follows from the inequalities $j'\ge j_2-p_3$, 
$j_2\ge j\ge j_3-(2p_3/3)$ and $q'\le q<p_3/3$ that $j'+1-q'>i_3$. Thus
$i_3\le j'+1-q'<j'+1\le j_3-p_3$, i.e. both symbols $w[j'+1-q']$ and $w[j'+1]$ 
are contained in $u'_3$. Since $u'_3=u''_3$, we obtain that the symbols
$w[j'+1+p_3]$ and $w[j'+1+p_3-q']$ contained in $u''_3$ are also different.
It follows from inequalities $j'\ge j_2-p_3$, $j_2\ge j\ge j_1-(2p_1/3)$ and
$q'\le q<p_3/3<p_1/3$ that $j'+1+p_3-q'>j_1-p_1$. On the other hand, since
$w[j'+1+p_3]$ is contained in $u''_3$, we have $j'+1+p_3\le j_3<j_1$. Thus
both symbols $w[j'+1+p_3]$ and $w[j'+1+p_3-q']$ are contained in $u''_1$.
Since $u'_1=u''_1$, we conclude that the symbols $w[j'+1+p_3-p_1]$ and 
$w[j'+1+p_3-q'-p_1]$ contained in $u'_1$ are also different. The inequality
$p_3<p_1$ implies that $j'+1+p_3-p_1\le j'$. On the other hand, since
$w[j'+1+p_3-q'-p_1]$ is contained in $u'_1$, we obtain
$$
j'+1+p_3-q'-p_1\ge i_1\ge 1_3-q\ge i'.
$$
Thus both symbols $w[j'+1+p_3-p_1]$ and $w[j'+1+p_3-q'-p_1]$ are contained 
in $\hat v'$ which contradicts the fact that $q'$ is a period of~$\hat v'$.
So this subcase is also impossible.

Subcase I.2.b. Let $j'\ge j_3-p_3$. In this case $u'_3$ is contained in $\hat v'$, 
so $q'$ is a period of $u'_3$. Since $|u'_3|=p_3>3q\ge 3q'$, using Proposition~\ref{repper},
it is easy to see that $q'$ has to be the minimal period of $u'_3$. Thus, $u'_3$
is a repetition with the minimal period~$q'$. So $u''_3$ is also a repetition with 
the minimal period~$q'$. Let $\hat v''\equiv w[i''..j'']$ be the maximal repetition 
containing~$u''_3$. If $\hat v'\equiv\hat v''$, then $u'_3u''_3$ is contained in $\hat v'$,
so $q'$ is the minimal period of $u'_3u''_3$. Applying Proposition~\ref{repper} to $u'_3u''_3$,
we obtain that in this case $q'$ is a divisor of $p_3$, so $q'$ is a period of $r_3$ 
which contradicts the fact that $p_3$ is the minimal period of $r_3$. Thus $\hat v'\not\equiv\hat v''$.
It is obvious that the repetitions $\hat v'$, $\hat v''$ are non-separated and cognate.
Taking into account $i'\le i_3-q$, $s_3=i_3$, and $j'\ge j_3-p_3$, we also have $i'<s_3\le j'$. 
Note that, obviously, $i_1<j_3-p_3$. Consider the factor $v'_1=w[i_1..j_3-p_3]$. 
The inequlities $i_3-q\le i_1$ and $p_1>p_2$ imply that $j_1-p_1>j_3-p_3$, 
so $v'_1$ is contained in $u'_1$. Therefore, $u'_1=u''_1$ implies $v'_1=v''_1$ where 
$v''_1\equiv w[i_1+p_1..j_3+p_1-p_3]$. It follows from $i'\le i_3-q\le i_1$ and 
$j'\ge j_3-p_3$ that $v'_1$ is also contained in $\hat v'$, so $q'$ is a period of $v'_1$.
Hence $q'$ is also a period of $v''_1$. From $j_1-p_1>j_3-p_3$ and $p_1>p_3$ we have
$i_1+p_1>i_3+p_3$ and $j_3+p_1-p_3>j_3$, so the overlap of $v''_1$ and $u''_3$ is
$w[i_1+p_1..j_3]$. The inequlities $j_3\ge j\ge j_1-(2p_1/3)$ imply that the length
of this overlap is no less than $p_1/3>q\ge q'$. Hence, using Proposition~\ref{unionper},
we obtain that $q'$ is the minimal period of $w[i_3+p_3..j_3+p_1-p_3]$. So $w[i_3+p_3..j_3+p_1-p_3]$
is contained in $\hat v''$. Thus 
$$
j''\ge j_3+p_1-p_3>j_3+p_2-p_3=j_3+q.
$$
Therefore, if $t_3\ge j_3+q$, then both numbers $q'$ and $p_3$ are periods of the factor 
$w[i_3+p_3..j_3+q]$ and, moreover, the length of this factor is $p_3+q$, i.e. is no less 
than $p_3+q'$. Hence, by the periodicity lemma, in this case $w[i_3+p_3..j_3+q]$ has the 
period $\gcd (q', p_3)$ which contradicts the primitivity of cyclic roots of~$r_3$. Thus, 
$t_3<j_3+q$, i.e. $t_3<j''$. On the other hand, we have, obviously, $t_3\ge i_3+p_3\ge i''$.
Recall also that $p_3>3q\ge 3q'$. Summing up the inequalities proved above, we obtain that
$r_3$ is generated by the repetitions $\hat v'$ and $\hat v''$, i.e. $r_3$ is a secondary
repetition which contradicts $r_3\in Rp (w)$. Thus, Case I is impossible.

Case II. Let $j_3<j_2$ and $i_3>i_2$. In this case we consider separately the three following 
subcases: $j_3-p_3>j_2-p_2$, $j_3-p_3=j_2-p_2$, and $j_3-p_3<j_2-p_2$.

Subcase II.1. Let $j_3-p_3>j_2-p_2$. Denote for convenience the root $u''_3$ by~$v$. Note that 
in this subcase $v$ is contained completely in $u''_2$. Thus, from $u'_2=u''_2$ and $u'_3=u''_3$
we obtain
\begin{equation}
v=w[i_3+p_3-p_2..j_3-p_2]=w[i_3..j_3-p_3].
\label{doublev1}
\end{equation}
So $q$ is a period of~$v$. Moreover, $|v|=p_3>3q$. Thus, $v$ is a repetition, and by Proposition~\ref{repper}
the minimal period $q'$ of this repetition is a divisor of~$q$. Denote by $v'$ the factor
$w[i_3+p_3-p_2..j_3-p_3]$. From~(\ref{doublev1}) we have that $v'$ is also a repetition with 
the minimal period $q'$. Let $s_3\neq i_3$, i.e. $r_3$ has at least one cyclic root to the left 
of $w[i_3]$. Then $v'$ is contained in $r_3$, so $v'$ has both periods $q'$ and $p_3$, and,
moreover, $|v'|=p_3+q\ge p_3+q'$. Therefore, by the periodicity lemma $v'$ has the period $\gcd (p_3, q')$
which contradicts the primitivity of cyclic roots of~$r_3$. Thus, $s_3=i_3$. Since $i_3>1$, it implies
$w[i_3-1]\neq w[j_3-p_3]$. Since $j_3-p_3>j_2-p_2$, the letter $w[j_3-p_3]$ is contained in $u''_2$,
so $w[j_3-p_3]=w[j_3-p_3-p_2]$. Hence $w[i_3-1]\neq w[j_3-p_3-p_2]=w[i_3-q-1]$. In this subcase we 
consider separately the two following subcases.

Subcase II.1.a. Let $i_1\le j_3-p_3-p_2$. From inequalities $j_3-(2p_3/3)\le j\le j_1$ and
$p_1\le 4p_3/3$ we have that $j_1-p_1\ge j_3-2p_3=i_3-1$. Thus
$$
i_1\le j_3-p_3-p_2<i_3-1\le j_1-p_1,
$$
i.e. both symbols $w[i_3-1]$ and $w[j_3-p_3-p_2]$ are contained in $u'_1$. So $w[i_3-1]=w[i_3+p_1-1]$
and $w[j_3-p_3-p_2]=w[j_3+p_1-p_3-p_2]$. Thus $w[i_3+p_1-1]\neq w[j_3+p_1-p_3-p_2]$. Using $p_1>p_2$, 
we obtain that $j_3+p_1-p_3-p_2\ge j_3+1-p_3=i_3+p_3$. On the other hand, the inequality
$p_1\le 4p_3/3$ implies $i_3+p_1-1<i_3+2p_3-1=j_3$. Thus we have that
$$
i_3+p_3\le j_3+p_1-p_3-p_2=i_3+p_1-q-1<i_3+p_1-1<j_3
$$
i.e. both letters $w[i_3+p_1-1]$ and $w[j_3+p_1-p_3-p_2]$ are contained in~$v$. Therefore,
since $j_3+p_1-p_3-p_2=i_3+p_1-q-1$, the relation $w[i_3+p_1-1]\neq w[j_3+p_1-p_3-p_2]$
contradicts the fact that $q$ is a period of~$v$. So this subcase is impossible. 

Subcase II.1.b. Let $i_1>j_3-p_3-p_2$. Consider the maximal repetitions $\hat v'\equiv w[i'..j']$
and $\hat v''\equiv w[i''..j'']$ containing respectively $v'$ and $v''$ with the minimal period~$q'$.
By the same way as in subcase I.2.b we can prove that $\hat v'\not\equiv\hat v''$.
Moreover, it is obvious that the repetitions $\hat v'$, $\hat v''$ are non-separated and cognate.
Denote by $v'_1$ the factor $w[i_1..j_3-p_3]$. Note that $v'_1$ is contained in $v'$,
so $q'$ is a period of $v'_1$. It follows from $i_1>j_3-p_3-p_2$ and $p_1>p_2$ that $j_1-p_1>j_3-p_3$. 
So $v'_1$ is contained in $u'_1$. Therefore, $u'_1=u''_1$ implies $v'_1=v''_1$ where
$v''_1\equiv w[i_1+p_1..j_3+p_1-p_3]$. So $q'$ is also a period of $v''_1$. Since
$j_1-(2p_1/3)\le j\le j_3$, we have $j_1-p_1\le j_3-(p_1/3)$. On the other hand,
from $p_1>p_3$ we have $j_3+p_1-p_3>j_3$. Thus, the length of the overlap $w[i_1+p_1..j_3]$
of $v$ and $v''_1$ is not less than $p_1/3$, i.e. is greater than~$q'$. Hence, by Proposition~\ref{unionper},
we obtain that $q'$ is the minimal period of $w[i_3+p_3..j_3+p_1-p_3]$. So $w[i_3+p_3..j_3+p_1-p_3]$
is contained in $\hat v''$. Then, by the same way as in subcase I.2.b we can show that
$i''\le t_3<j''$. We have also that 
$$
i'\le i_3+p_3-p_2=i_3-q<i_3=s_3\le j_3-p_3\le j'.
$$
Thus, as in subcase I.2.b, we obtain that $r_3$ is generated by $\hat v'$ and $\hat v''$, i.e. $r_3$ is 
a secondary repetition which contradicts $r_3\in Rp (w)$. So subcase II.1 is impossible.

Subcase II.2. Let $j_3-p_3=j_2-p_2$. Then from $u'_2=u''_2$ and $u'_3=u''_3$ we obtain that $q$ is a period 
of $u'_2$ and $u''_2$. Moreover, taking into account $|u'_2|=|u''_2|=p_2>p_3>3q$ and Proposition~\ref{repper},
we have that $u'_2$ and $u''_2$ are repetitions, and the minimal period $q'$ of these repetitions is a divisor 
of~$q$. Consider again the maximal repetitions $\hat v'\equiv w[i'..j']$ and $\hat v''\equiv w[i''..j'']$ 
with the minimal period~$q'$ containing respectively $u'_2$ and $u''_2$. By the same way as in subcase I.2.b 
we can prove that $\hat v'\not\equiv\hat v''$. If $s_3\neq i_3$ then $u'_2$ is contained completely in $r_3$,
so $u'_2$ has both periods $q$ and $p_3$. Since $|u'_2|=p_2=p_3+q$, using the periodicity lemma, we obtain
in this case that $u'_2$ has the period $\gcd (p_3, q)$, so $u'_3$ has also the period $\gcd (p_3, q)$
which contradicts the primitivity of cyclic roots of~$r_3$. Thus, $s_3=i_3$. Therefore,
$$
i'\le i_2<i_3=s_3\le j_3-p_3=j_2-p_2\le j'.
$$
If $t_3\ge j_2$ then $u''_2$ is contained completely in $r_3$, so in this case we can also obtain 
a contradiction to the primitivity of cyclic roots of~$r_3$. Hence 
$$
i''\le i_2+p_2=i_3+p_3\le j_3\le t_3<j_2\le j''.
$$
It is also obvious that $\hat v'$, $\hat v''$ are non-separated and cognate. Thus, taking into account 
the inequalities proved above, we obtain in this subcase that $r_3$ is generated by $\hat v'$ and $\hat v''$ 
which contradicts $r_3\in Rp (w)$. 

Subcase II.3. Let $j_3-p_3<j_2-p_2$. Denote by $v'$ and $v''$ the factors $w[i_2..j_3-p_3]$ and $w[i_3+p_3..j_3+q]$
respectively. From $u'_2=u''_2$ and $u'_3=u''_3$ we have 
$$
w[i_2+q..j_3-p_3]=w[i_2+p_2..j_3]=w[i_2..j_3-p_2],
$$
so $q$ is a period of~$v'$. Since $u'_3$ is contained in $u'_2$, by the same way we have
$$
u''_3\equiv w[i_3+p_3..j_3]=u'_3=w[i_3+p_2..j_3+q],
$$
so $q$ is also a period of~$v''$. Since $|v'|,|v''|>p_3>3q$, we obtain that $v'$, $v''$ are cognate
repetitions, and the minimal period $q'$ of these repetitions is a divisor of~$q$.

First we prove that $s_3=i_3$. Assume that $s_3\neq i_3$, i.e $s_3\le i_3-p_3$. If $s_2\neq i_2$,
i.e. $s_2\le i_2-p_2<i_3-p_3$, then the factor $w[i_3-p_3..j_3]$ of length $3p_3$ has both periods $p_3$ and $p_2$.
Since $p_2+p_3<7p_3/3<3p_3$, by the periodicity lemma we obtain in this case that this factor has
the period $\gcd (p_3, p_2)$ which contradicts the primitivity of cyclic roots of~$r_2$. Thus, $s_2=i_2$.
Let $\hat v'\equiv w[i'..j']$ be the maximal repetition containing the repetition~$v'$, and 
$\hat v''$ be the maximal repetition containing the repetition~$v''$. By the same way as in 
subcase I.2.b we can prove that $\hat v'\not\equiv\hat v''$. We consider separately 
the three following subcases.

Subcase II.3.a. Let $i_1<i_2$. Then $i_2>1$, so $s_2=i_2$ implies that $w[i_2-1]\neq w[j_2-p_2]$.
It follows from $j_3-p_3<j_2-p_2$ that the letter $w[j_2-p_2]$ is contained in $u''_3$, so
$w[j_2-p_2]=w[j_2-p_2-p_3]=w[i_2+q-1]$. Thus, $w[i_2-1]\neq w[i_2+q-1]$. Note that 
$$
p_1\le\frac{4}{3}p_3<p_3+\frac{1}{3}p_2=\frac{4}{3}p_2-q.
$$
Using this estimation together with $j_1\ge j\ge j_2-(2p_2/3)$, we obtain
$j_1-p_1>j_2+q-2p_2=i_2+q-1$. Thus, $i_1\le i_2-1<i_2+q-1<j_1-p_1$, i.e.
both letters $w[i_2-1]$, $w[i_2+q-1]$ are contained in $u'_1$. Therefore,
$w[i_2-1]=w[i_2+p_1-1]$ and $w[i_2+q-1]=w[i_2+p_1+q-1]$. Hence
$w[i_2+p_1-1]\neq w[i_2+p_1+q-1]$. Using $p_1>p_2$, we have $i_2+p_1-1\ge i_2+p_2$,
so $i_2+p_1-1>i_3+p_3$. On the other hand, using $p_1\le 4p_3/3<2p_3$, we have
$$
i_2+p_1+q-1<i_2+2p_3+q-1<i_3+2p_3+q-1=j_3+q.
$$
Thus, both letters $w[i_2+p_1-1]$, $w[i_2+p_1+q-1]$ are contained in $v''$.
Therefore, $w[i_2+p_1-1]\neq w[i_2+p_1+q-1]$ contradicts the fact that $q$ 
is a period of~$v''$. So this subcase is impossible.

Subcase II.3.b. Let $i_1\ge i_2$ and $j'<j_2-p_2$. Since $\hat v'$ is maximal,
we have $w[j'+1]\neq w[j'+1-q']$. From $j_3-p_3\le j'<j_2-p_2$ and $q'\le q<p_3$
we have also that $i_3<j'+1-q'<j'+1\le j_2-p_2$, i.e. both letters $w[j'+1]$, $w[j'+1-q']$
are contained in $u'_2$. So $w[j'+1]=w[j'+p_2+1]$ and $w[j'+1-q']=w[j'+p_2+1-q']$.
Thus $w[j'+p_2+1]\neq w[j'+p_2+1-q']$. Note that in this subcase $j_1>j_2$ and
$j'+p_2<j_2$, so $j'+p_2+1<j_1$. On the other hand, we have $j'\ge j_3-p_3$, so
$j'+p_2+1>j_3+p_2-p_3=j_3+q\ge j_3+q'$. Hence $j'+p_2+1-q'>j_3$. Taking into account
$j_3\ge j\ge j_1-(2p_1/3)>j_1-p_1$, we obtain
$$
j_1-p_1<j'+p_2+1-q'<j'+p_2+1<j_1.
$$
Thus, both letters $w[j'+p_2+1-q']$ and $w[j'+p_2+1]$ are contained in $u''_1$.
Hence $w[j'+p_2+1-q']=w[j'+p_2+1-q'-p_1]$ and $w[j'+p_2+1]=w[j'+p_2+1-p_1]$.
So $w[j'+p_2+1-q'-p_1]\neq w[j'+p_2+1-p_1]$. Since $p_2<p_1$, we have
$j'+p_2+1-p_1\le j'$. On the other hand, since $w[j'+p_2+1-q'-p_1]$ is contained
in $u'_1$, we have also $j'+p_2+1-q'-p_1\ge i_1\ge i_2\ge i'$. Thus,
both letters $w[j'+p_2+1-q'-p_1]$ and $w[j'+p_2+1-p_1]$ are contained
in $\hat v'$. So $w[j'+p_2+1-q'-p_1]\neq w[j'+p_2+1-p_1]$ contradicts 
the fact that $q'$ is a period of~$\hat v'$. Therefore, this subcase is
also impossible.

Subcase II.3.c. Let $j'\ge j_2-p_2$. Then $u'_2$ is contained completely
in $\hat v'$, so $q'$ is a period of $u'_2$. It follows from $j_3-p_3<j_2-p_2$
and $p_2<4p_3/3<2p_3$ that $i_3-p_3<i_2$, so $u'_2$ is contained completely
in $r_3$, i.e. $p_3$ is also a period of $u'_2$. Moreover, $|u'_2|=p_2=p_3+q\ge p_3+q'$.
Therefore, by the periodicity lemma this factor has the period $\gcd (p_3, q')$ which 
contradicts the primitivity of the root $u'_3$ contained in $u'_2$.

Since all the considered subcases are impossible, we conclude that $s_3=i_3$. Then,
analogously to subcase II.2, one can prove that $r_3$ is generated by $\hat v'$ and $\hat v''$ 
which contradicts $r_3\in Rp (w)$. Thus, Case I is also impossible.

Case III. Let $i_3\le i_2$. In this case we consider separately the subcases $s_2=i_2$
and $s_2\neq i_2$.

Subcase III.1. Let $s_2=i_2$. Denote by $v$ the overlap $w[i_2..j_3-p_3]$ of $u'_2$ and $u'_3$.
It follows from $j_3\ge j\ge j_2-(2p_2/3)$ and $q<p_2/3$ that $|v|\ge (p_2/3)+q>2q$. Since
$u'_2=u''_2$ and $u'_3=u''_3$, we have
\begin{equation}
v=w[i_2+p_2..j_3+q]=w[i_2+p_3..j_3],
\label{doublev2}
\end{equation}
so $q$ is a period of~$v$. Thus, $v$ is a repetition, and by Proposition~\ref{repper}
the minimal period~$q'$ of this repetition is a divisor of~$q$. Therefore, using
again~(\ref{doublev2}), we obtain that $w[i_2+p_3..j_3+q]$ is also a repetition with
the minimal period~$q'$. We denote this repetition by~$v''$ and consider separately 
the subcases $i_1<i_2$ and $i_1\ge i_2$.

Subcase III.1.a. Let $i_1<i_2$. Note that in this subcase $i_2>1$, so
$s_2=i_2$ implies that $w[i_2-1]\neq w[j_2-p_2]$. It follows from $i_3\le i_2$,
$p_3<p_2$ and $j_3\ge j\ge j_2-(2p_2/3)$ that $i_3+p_3\le j_2-p_2<j_3$, i.e.
the letter $w[j_2-p_2]$ is contained in $u''_3$. So $w[j_2-p_2]=w[j_2-p_2-p_3]=w[i_2+q-1]$.
Thus $w[i_2-1]\neq w[i_2+q-1]$. Note that
$$
\frac{4}{3}p_2-q=\frac{1}{3}p_2+p_3>\frac{4}{3}p_3\ge p_1.
$$
Therefore, $j_1\ge j\ge j_2-(2p_2/3)$ implies that $j_1-p_1>j_2+q-2p_2=i_2+q-1$.
On the other hand, we have $i_1\le i_2-1$. Thus, both letters $w[i_2-1]$ and $w[i_2+q-1]$
are contained in $u'_1$. So $w[i_2-1]=w[i_2+p_1-1]$ and $w[i_2+q-1]=w[i_2+p_1+q-1]$.
Therefore, $w[i_2+p_1-1]\neq w[i_2+p_1+q-1]$. From $p_1>p_2$ we have $i_2+p_1-1\ge i_2+p_2>i_2+p_3$.
On the other hand, using $p_1<4p_2/3$ and $j_3\ge j\ge j_2-(2p_2/3)$, we obtain
$$
i_2+p_1+q-1=j_2+p_1+q-2p_2<j_2+q-\frac{2}{3}p_2\le j_3+q.
$$
Thus, both letters $w[i_2+p_1-1]$ and $w[i_2+p_1+q-1]$ are contained in~$v''$.
Therefore, since $q'$ is a divisor of~$q$, the inequality $w[i_2+p_1-1]\neq w[i_2+p_1+q-1]$
contradicts the fact that $q'$ is a period of~$v''$. So this subcase is impossible.

Subcase III.1.b. Let $i_1\ge i_2$. Denote by $\hat v'\equiv w[i'..j']$ the maximal repetition
contaning the repetition~$v$ and by $\hat v''\equiv w[i''..j'']$ the maximal repetition
contaning the repetition~$v''$. Let $\hat v'\equiv\hat v''$. Then $j'\ge j_3+q$, so $u'_2$ is 
contained completely in $\hat v'$. Therefore, $q'$ is the minimal period of $u'_2$, so
$q'$ is also the minimal period of $u''_2$. Since $u''_2$ is overlapped with $\hat v'$
by at least $|v|$ letters where $|v|>2q>q'$, by Proposition~\ref{unionper} we obtain in this case 
that $q'$ is the minimal period of $u'_2u''_2$. So, by Proposition~\ref{repper}, $q'$ is a divisor 
of $p_2$ which contradicts the primitivity of cyclic roots of~$r_2$. Thus, $\hat v'\not\equiv\hat v''$.
By Proposition~\ref{overlap1}, $\hat v'$ and $\hat v''$ can not be overlapped by greater than or equal 
to $q'$ letters, so $i''\le i_2+p_3$ and $q'\le q$ imply that $j'<j_2-p_2$. Since the repetition 
$\hat v'$ is maximal, we have $w[j'+1-q']\neq w[j'+1]$. It follows from $j'+1\ge i_2+|v|>i_2+2q\ge i_2+2q'$
and $j'<j_2-p_2$ that $i_2<j'+1-q'<j'+1\le j_2-p_2$, i.e. both letters $w[j'+1-q']$ and $w[j'+1]$
are contained in $u'_2$. Therefore, $w[j'+1-q']=w[j'+p_2+1-q']$ and $w[j'+1]=w[j'+p_2+1]$.
Thus, $w[j'+p_2+1-q']\neq w[j'+p_2+1]$. From $i_1\ge i_2$ we obtain $j_1>j_2$. Therefore,
since $w[j'+p_2+1]$ is contained in $u''_2$, we have $j'+p_2+1<j_1$. On the other hand, 
$j'\ge j_3-p_3$ implies that $j'+p_2+1>j_3+q$, so 
$$
j'+p_2+1-q'>j_3+q-q'\ge j_3\ge j\ge j_1-\frac{2p_1}{3}>j_1-p_1.
$$
Thus, both letters $w[j'+p_2+1-q']$ and $w[j'+p_2+1]$ are contained in $u''_1$.
Therefore, $w[j'+p_2+1-q']=w[j'+p_2+1-p_1-q']$ and $w[j'+p_2+1]=w[j'+p_2+1-p_1]$.
Thus, $w[j'+p_2+1-p_1-q']\neq w[j'+p_2+1-p_1]$. It follows from $p_1>p_2$ that
$j'+p_2+1-p_1\le j'$. On the other hand, since $w[j'+p_2+1-p_1-q']$ is contained
in $u'_1$, we have $j'+p_2+1-p_1-q'\ge i_1$, so $j'+p_2+1-p_1-q'\ge i_2\ge i'$.
Thus, both letters $w[j'+p_2+1-p_1-q']$ and $w[j'+p_2+1-p_1]$ are contained in $\hat v'$.
Hence $w[j'+p_2+1-p_1-q']\neq w[j'+p_2+1-p_1]$ contradicts the fact that $q'$ is 
a period of~$\hat v'$. Thus, subcase III.1 is impossible.

Subcase III.2. Let $s_2\neq i_2$, i.e. $s_2\le i_2-p_2$. Then, analogously to case I, 
one can prove that $s_3=i_3$. Denote respectively by $u'$ and $u''$ the factors
$w[i_3-q..j_3-p_3]$ and $w[i_3+p_3..j_3+q]$. It is easy to see that $j_3\ge j\ge 
j_2-(2p_2/3)>j_2-p_2$ implies $i_3-q>i_2-p_2$, i.e. $i_3-q>s_2$, and $i_3\le i_2$
implies $j_3+q<j_2$, i.e. $j_3+q<t_2$. Thus, $u'$ and $u''$ are contained completely
in $r_2$, so $u'$ and $u''$ are cyclic roots of $r_2$. Note that if one considers respectively
$u'$ and $u''$ instead of $u'_2$ and $u''_2$, this subcase is identical to subcase II.2.
Hence, by the same way as in subcase II.2, we can prove that in this subcase the repetition $r_3$
is secondary. This contradiction to $r_3\in Rp (w)$ completes the proof of Lemma~\ref{baselemma}.

Further  we assign to each point $(p, j)$ the weight $\rho (p, j)=1/p^2$, and for any finite set~$A$ 
of points we define 
$$
\rho (A)=\sum_{(p, j)\in A} \rho (p, j)=\sum_{(p, j)\in A}\frac{1}{p^2}.
$$
Let $\lambda$ be a positive integer. By ${\cal E}_{\lambda}(w)$ (${\cal E}'_{\lambda}(w)$) we denote 
the set of all points $(p, j)$ from ${\cal E}(w)$ (${\cal E}'(w)$) such that $p\ge\lambda$.
Using Lemma~\ref{baselemma}, we prove the following 
\begin{corollary}
$|{\cal E}'_{\lambda}(w)|=O\left(\frac{n}{\lambda}\right)$.
\label{elambda}
\end{corollary}

{\bf Proof.} It is obvious that for any point $(p, j)$ from ${\cal E}'_{\lambda}(w)$ the set
$V(p, j)$ is contained in ${\cal E}_{\lambda}(w)$. On the other hand, by Lemma~\ref{baselemma},
each point of ${\cal E}_{\lambda}(w)$ can not be covered by more than two points of ${\cal E}'_{\lambda}(w)$.
Therefore,
$$
\sum_{(p, j)\in {\cal E}'_{\lambda}(w)} \rho (V(p, j))\le 2\rho ({\cal E}_{\lambda}(w))=
2\left(\sum_{(p, j)\in {\cal E}_{\lambda}(w)}\frac{1}{p^2}\right)=
2n\left(\sum_{\lambda\le p\le 2n/3}\frac{1}{p^2}\right).
$$
Using the evident inequality $\frac{1}{p^2}<\int_{p-\frac{1}{2}}^{p+\frac{1}{2}}\frac{1}{x^2}\,dx$,
we estimate
$$
\sum_{\lambda\le p\le 2n/3}\frac{1}{p^2}<\sum_{p=\lambda}^{\infty}\frac{1}{p^2}<
\int_{\lambda-\frac{1}{2}}^{\infty}\frac{1}{x^2}\,dx =\frac{1}{\lambda-\frac{1}{2}}.
$$
Thus,
$$
\sum_{(p, j)\in {\cal E}'_{\lambda}(w)} \rho (V(p, j))<\frac{2n}{\lambda-\frac{1}{2}}.
$$
On the other hand, for any point $(p, j)$ we can also estimate $\rho (V(p, j))$:
\begin{eqnarray*}
\rho (V(p, j))&>&\frac{2p}{3}\sum_{p\le i\le 4p/3}\frac{1}{i^2}>
\frac{2p}{3}\int_p^{\lfloor 4p/3\rfloor +1}\frac{1}{x^2}\,dx\\
&>&\frac{2p}{3}\int_p^{4p/3}\frac{1}{x^2}\,dx=\frac{2p}{3}\cdot\frac{1}{4p}=\frac{1}{6}.
\end{eqnarray*}
So $\sum_{(p, j)\in {\cal E}'_{\lambda}(w)} \rho(V(p, j))>\frac{1}{6} |{\cal E}'_{\lambda}(w)|$.
Therefore,
$$
|{\cal E}'_{\lambda}(w)|<6\left(\sum_{(p, j)\in {\cal E}'_{\lambda}(w)} \rho(V(p, j))\right)
<\frac{12n}{\lambda-\frac{1}{2}}=O\left(\frac{n}{\lambda}\right).
$$

Let $Rp_{\lambda}(w)$ be the set of all repetitions from $Rp (w)$ with the minimal period greater
than or equal to~$\lambda$. It is obvious that ${\cal E}'_{\lambda}(w)=
\bigcup_{r\in Rp_{\lambda}(w)} {\cal P}(r)$. Therefore, since all the sets ${\cal P}(r)$ for 
$r\in Rp_{\lambda}(w)$ are non-empty and pairwise disjoint by Proposition~\ref{notinter}, 
the bound $|Rp_{\lambda}(w)|\le ||{\cal E}'_{\lambda}(w)|$ takes place. Thus, 
Corollary~\ref{elambda} implies

\begin{theorem}
$|Rp_{\lambda}(w)|=O\left(\frac{n}{\lambda}\right)$.
\label{rlambda}
\end{theorem}

Corollary~\ref{elambda} allows actually to strengthen this result. Let $Rs_{\lambda}(w)$ 
be the set of all secondary repetitions generated by repetitions from $Rp_{\lambda}(w)$.
Denote by $exp_{\lambda}(w)$ the sum $\sum_{r\in Rp_{\lambda}(w)} e(r)$ of the exponents
of all repetitions from $Rp_{\lambda}(w)$ and by $exs_{\lambda}(w)$ the sum 
$\sum_{r\in Rs_{\lambda}(w)} e(r)$ of the exponents of all repetitions from $Rs_{\lambda}(w)$.
Then we have

\begin{theorem}
$exp_{\lambda}(w)+exs_{\lambda}(w)=O\left(\frac{n}{\lambda}\right)$.
\label{exlambda}
\end{theorem}

{\bf Proof.} By Lemma~\ref{paruniq} each repetition $r$ from $Rs_{\lambda}(w)$ can be 
corresponded to the repetition from $Rs_{\lambda}(w)$ which generates~$r$ from left.  
Hence $|Rs_{\lambda}(w)|\le \sum_{r\in Rp_{\lambda}(w)} e(r)-2$, due to Corollary~\ref{descr}. 
From the proof of Corollary~\ref{secbyprim} we can also conclude that the exponent of any 
secondary repetition is less than~3. Therefore, 
$$
exs_{\lambda}(w)<3\cdot |Rs_{\lambda}(w)|\le \sum_{r\in Rp_{\lambda}(w)} 3e(r)-6.
$$
Thus,
$$
exp_{\lambda}(w)+exs_{\lambda}(w)<\sum_{r\in Rp_{\lambda}(w)} 4e(r)-6=
4\cdot\left(\sum_{r\in Rp_{\lambda}(w)} e(r)-\frac{3}{2}\right).
$$
Using $|{\cal P}(r)|=\lfloor e(r)-1\rfloor$, we can estimate $e(r)-\frac{3}{2}<\frac{3}{2} |{\cal P}(r)|$,
so $\sum_{r\in Rp_{\lambda}(w)} e(r)-\frac{3}{2}<\frac{3}{2} |{\cal E}'_{\lambda}(w)|$.
Therefore, $exp_{\lambda}(w)+exs_{\lambda}(w)<6\cdot |{\cal E}'_{\lambda}(w)|$.
Hence Theorem~\ref{exlambda} follows immediately from Corollary~\ref{elambda}.

Let $clp_{\lambda}(w, i)$ where $i=1, 2,\ldots ,n$ be the number of repetitions from $Rp_{\lambda}(w)$
which contain the letter $w[i]$, and $clp_{\lambda}(w)=\max_i clp_{\lambda}(w, i)$. The value $clp_{\lambda}(w)$
can be also estimated by Lemma~\ref{baselemma}.

\begin{theorem}
$clp_{\lambda}(w)=O(\log \frac{n}{\lambda})$.
\label{clambda}
\end{theorem}

{\bf Proof.} Consider the number~$i$ sach that $clp_{\lambda}(w)=clp_{\lambda}(w, i)$. Denote by $R'$
the set of all repetitions from $Rp_{\lambda}(w)$ which contain $w[i]$. We correspond each repetition~$r$
from $R'$ to some point $(p(r), j_r)$ of ${\cal P}(r)$ in the following way. If in ${\cal P}(r)$ there exists
at least one point $(p(r), j)$ such that $j\ge i$ then $j_r$ is the minimal number~$j$ such that $(p(r), j)\in {\cal P}(r)$
and $j\ge i$. Othervise $j_r$ is the maximal number~$j$ such that $(p(r), j)\in {\cal P}(r)$. It is easy to
note that in this case $i-p(r)<j_r<i+2p(r)$. Therefore, for any repetition~$r$ from~$R'$ we have that $V(p(r), j_r)$
is contained completely in the set of all points $(p, j)$ such that
$$
\lambda\le p\le\frac{2n}{3},\qquad i-\frac{5p}{3}<j<i+2p.
$$
Denote this set by $\Omega$. By Proposition~\ref{notinter} different repetitions from $R'$ correspond to
different points, and by Lemma~\ref{baselemma} each point of $W$ can not be covered by more than 
two different points corresponding to repetitions from $R'$. Thus,
$$
\sum_{r\in R'} \rho (V(p(r), j_r))\le 2\rho (\Omega).
$$
Using the evident inequality $\frac{1}{p}<\int_{p-\frac{1}{2}}^{p+\frac{1}{2}}\frac{1}{x}\,dx$,
we estimate $\rho (\Omega)$:
\begin{eqnarray*}
\rho (\Omega)&=&\sum_{p=\lambda}^{\lfloor 2n/3\rfloor}\sum_{i-(5p/3)<j<i+2p} \frac{1}{p^2}<
\sum_{p=\lambda}^{\lfloor 2n/3\rfloor} \frac{11p}{3}\cdot\frac{1}{p^2}
=\frac{11}{3}\sum_{p=\lambda}^{\lfloor 2n/3\rfloor} \frac{1}{p}\\
&<&\frac{11}{3}\int_{\lambda-\frac{1}{2}}^{\lfloor 2n/3\rfloor +\frac{1}{2}}\frac{1}{x}\,dx 
<\frac{11}{3}\int_{\lambda-\frac{1}{2}}^n\frac{1}{x}\,dx = \frac{11}{3}\ln\frac{n}{\lambda-\frac{1}{2}}.  
\end{eqnarray*}
On the other hand, it is shown in the proof of Corollary~\ref{elambda} that $\rho (V(p, j))>1/6$
for any point $(p, j)$. So $\sum_{r\in R'} \rho (V(p(r), j_r))>|R'|/6$. Thus,
$$
|R'|<6\sum_{r\in R'} \rho (V(p(r), j_r))\le 12\rho (\Omega)<44\ln\frac{n}{\lambda-\frac{1}{2}}
=O(\log \frac{n}{\lambda}).
$$
Since $|R'|=clp_{\lambda}(w)$, we conclude $clp_{\lambda}(w)=O(\log \frac{n}{\lambda})$.

Thus, unlike the case of all maximal repetitions, only a logarithmic number of primary repetitions 
in a word can contain the same letter.

\section{Conclusion}

In the paper we define secondary repetitions as generated repetitions~$r$ 
satisfying the condition $p(r)\ge 3p$ where $p$ is the minimal period of
the generating repetitions. At the same time we suppose that the factor~3
in this condition is ``conventional'', i.e. we conjecture that for any
natural~$k\ge 3$ after replacing this condition by $p(r)\ge kp$ 
Theorems \ref{rlambda}, \ref{exlambda} and~\ref{clambda} will remain true.

In the introduction we give an example of word which has many secondary
repetitions. However, the total number of runs in this word is relatively
small in comparison with the maximum possible number of runs in a word.
This observations allows to make the conjecture that the words with 
the maximum possible number of runs have no secondary repetitions,
i.e. $mrn(n)$ coincides with the maximum possible number of primary 
repetitions in words of length~$n$.

\section*{Acknowledgments}

This work is partially supported by Russian Foundation for Fundamental Research 
(Grant 11-01-00508).

\end{document}